\documentclass[11pt,twoside]{article}
\usepackage{./asp2014}

\aspSuppressVolSlug
\resetcounters

\begin{document}

\title{Learning from history: Adaptive calibration of `tilting spine' fiber positioners}
\author{James~Gilbert$^1$ and Gavin~Dalton$^{1,2}$
\affil{$^1$Department of Physics, University of Oxford, Oxford, UK}
\affil{$^2$RALSpace, STFC Rutherford Appleton Laboratory, Didcot, UK}}

\paperauthor{James Gilbert}{james.gilbert@physics.ox.ac.uk}{0000-0001-5065-2101}{University of Oxford}{Department of Physics}{Oxford}{Oxfordshire}{OX1 3RH}{UK}
\paperauthor{Gavin Dalton}{gavin.dalton@physics.ox.ac.uk}{}{University of Oxford}{Department of Physics}{Oxford}{Oxfordshire}{OX1 3RH}{UK}

\begin{abstract}
This paper discusses a new approach for determining the calibration parameters of independently-actuated optical fibers in multi-object astronomical fiber positioning systems.  This work comes from the development of a new type of piezoelectric motor intended to enhance the `tilting spine' fiber positioning technology originally created by the Australian Astronomical Observatory.  Testing has shown that the motor's performance can vary depending on the fiber's location within its accessible field, meaning that an individual fiber is difficult calibrate with a one-time routine.  Better performance has resulted from constantly updating calibration parameters based on the observed movements of the fiber during normal closed-loop positioning.  Over time, location-specific historical data is amassed that can be used to better predict the results of a future fiber movement.  This is similar to a technique previously proposed by the Australian Astronomical Observatory, but with the addition of location-specific learning.  Results from a prototype system are presented, showing a significant reduction in overall positioning error when using this new approach.
\end{abstract}

\section{Introduction}

An increasing number of astronomical multi-object spectrographs require fast and accurate automated positioning of many optical fibers.  The Australian Astronomical Observatory's (AAO's) `tilting spine' technology is a simple and modular solution for such instruments, featuring independently-actuated fibers that can all move simultaneously \citep{echidna_advances_2014}.  Tilting spine technology first appeared in the AAO's FMOS Echidna fiber positioner for Subaru \citep{echidna_2008} and will soon be a part of the European Southern Observatory's 4MOST instrument on Vista \citep{4most_2014}.

\begin{figure}[ht!]
    \centering
    \includegraphics[width=13.5cm]{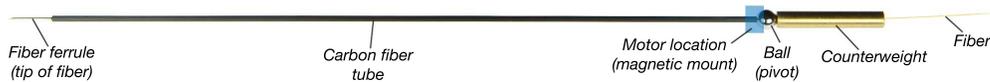}
    \caption{An example of a single-fiber tilting spine. The fiber is held within a rigid tube, which passes through a solid metal ball. A piezoelectric stick-slip motor tilts the spine by small discrete steps in one of several available directions.}
    \label{fig:spine}
\end{figure}

A single tilting spine is shown in Figure~\ref{fig:spine}.  It comprises a carbon fiber tube with a steel ball that sits in a magnetic mount driven by a piezo actuator.  This forms the basis of a stick-slip motor, able to tilt the spine and move the fiber tip in discrete steps of a few microns.  Many spines are packed together to form a high-density bed of fibers.

\subsection{Piezoelectric motor calibration}

Piezoelectric stick-slip motors offer simplicity at the cost of consistency between devices.  Tilting spine motors are no exception; they work by slightly tilting the spine's magnetic mount, before quickly returning to their initial state with an acceleration high enough to leave the spine in the displaced position.  This yields discrete angular stepping of the spine, hence moving the tip of the fiber.

\begin{figure}[ht!]
    \centering
    \includegraphics[height=1.5in]{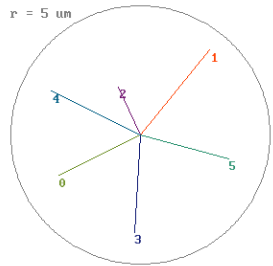}
    \caption{Calibration vectors for a tilting spine motor with six selectable movement directions. Most step magnitudes are approximately 4~\micron.}
    \label{fig:vectors}
\end{figure}

Tilting spine motors can move a spine in different directions, and so a fiber can be repositioned at any point in its patrol area with a combination of moves, each being of an appropriate number of motor steps.  It follows that the vector of a single step must be known before calculating the best move combination.  This is achieved with a motor calibration.  Figure~\ref{fig:vectors} shows a graphical representation of calibration data for a new design of tilting spine motor with six selectable movement directions.

The dependence of stick-slip motors on tribological factors means that their behavior can change over time, necessitating regular recalibration.  In \citet{echidna_advances_2014}, the AAO suggests remedying this by updating calibration data during normal positioning.  Since the system has a closed control loop with fiber position feedback, this is trivial to implement.  The AAO reports that this has been simulated with positive outcomes.

In investigating a new and improved piezoelectric motor design for tilting spines at the University of Oxford, it has become apparent that positioning performance can vary not only with time, but also as a function of target position.  This paper reports on laboratory testing of a simple yet effective adaptive calibration scheme that accounts for spatial as well as temporal changes in motor performance.

\section{Adaptive calibration approach}

\begin{figure}[ht!]
    \centering
    \includegraphics[height=6cm]{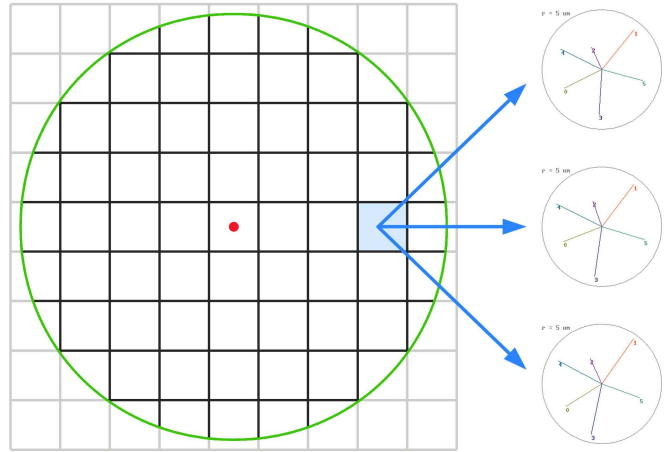}
    \caption{The patrol area of a spine is a circle (green line) of a certain radius (e.g.\ 5 mm) from the home position of the fiber (red dot). This area is split into an arbitrary number of `calibration cells', each of which has its own record of motor calibration vectors that were inferred from past moves.}
    \label{fig:calCells}
\end{figure}

The principle of the adaptive calibration scheme is simple.  First, the patrol area of each fiber is divided into a regular grid of virtual `calibration cells', where each cell holds its own calibration data.  Figure~\ref{fig:calCells} shows an example with 69 cells.  Whenever a spine is operating within a particular cell, its positioning errors are used to correct the motor calibration data.  This is added to a first-in-first-out (FIFO) stack of recent moves.

Before any move, the local calibration cell is queried and a set of historical calibration vectors is returned.  Median step sizes are then calculated, to exclude outlying or erroneous data.  This is similar to the AAO's proposed method for updating a spine's global calibration \citep{echidna_advances_2014}.

The number of calibration cells has so far been chosen empirically.  Too few cells will fail to characterize the spatial variation in motor behavior, whereas too many will lead to data becoming stale before it is used.  More cells also require more memory, scaling linearly.  For the example in Figure~\ref{fig:calCells}, in a 2500-fiber instrument (as is 4MOST), the total required memory would be 50~MB with double-precision numbers.

\section{Prototyping and testing}

\begin{figure}[ht!]
    \centering
    \includegraphics[height=6cm]{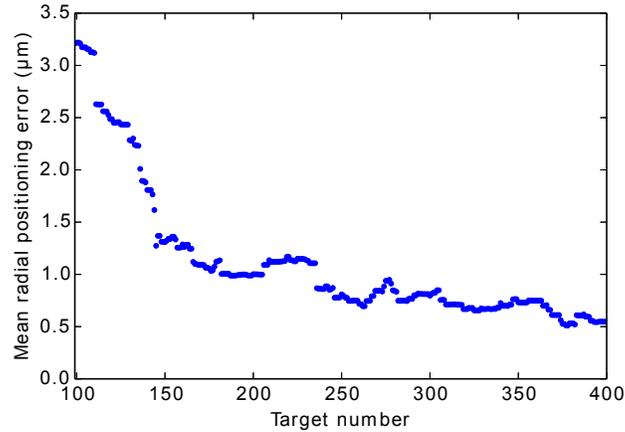}
    \caption{Running mean of positioning error (sample size 100) for 400 positioning cycles with a prototype spine. The error falls quickly at first, as the cells are updated for the first time, then reduces steadily until hitting the positioner's limit.}
    \label{fig:plot_err_history}
\end{figure}

Laboratory testing of a new type of tilting spine motor has shown a significant reduction of overall positioning error with when adaptive calibration is enabled.  The system's learning is evident in Figure~\ref{fig:plot_err_history}, with the average positioning error falling abruptly as the grid is updated for the first time.  Figure~\ref{fig:plot_calGrids} reveals a definite pattern in the spatial variation of spine behavior.

At the time of writing, adaptive calibration has been enabled for over 300~hours of continuous testing.  Motor performance has changed substantially throughout this time as a result of actuator burn-in and mechanical wear.  Thanks to the adaptive calibration scheme, positioning errors remain acceptable.

\articlefiguretwo{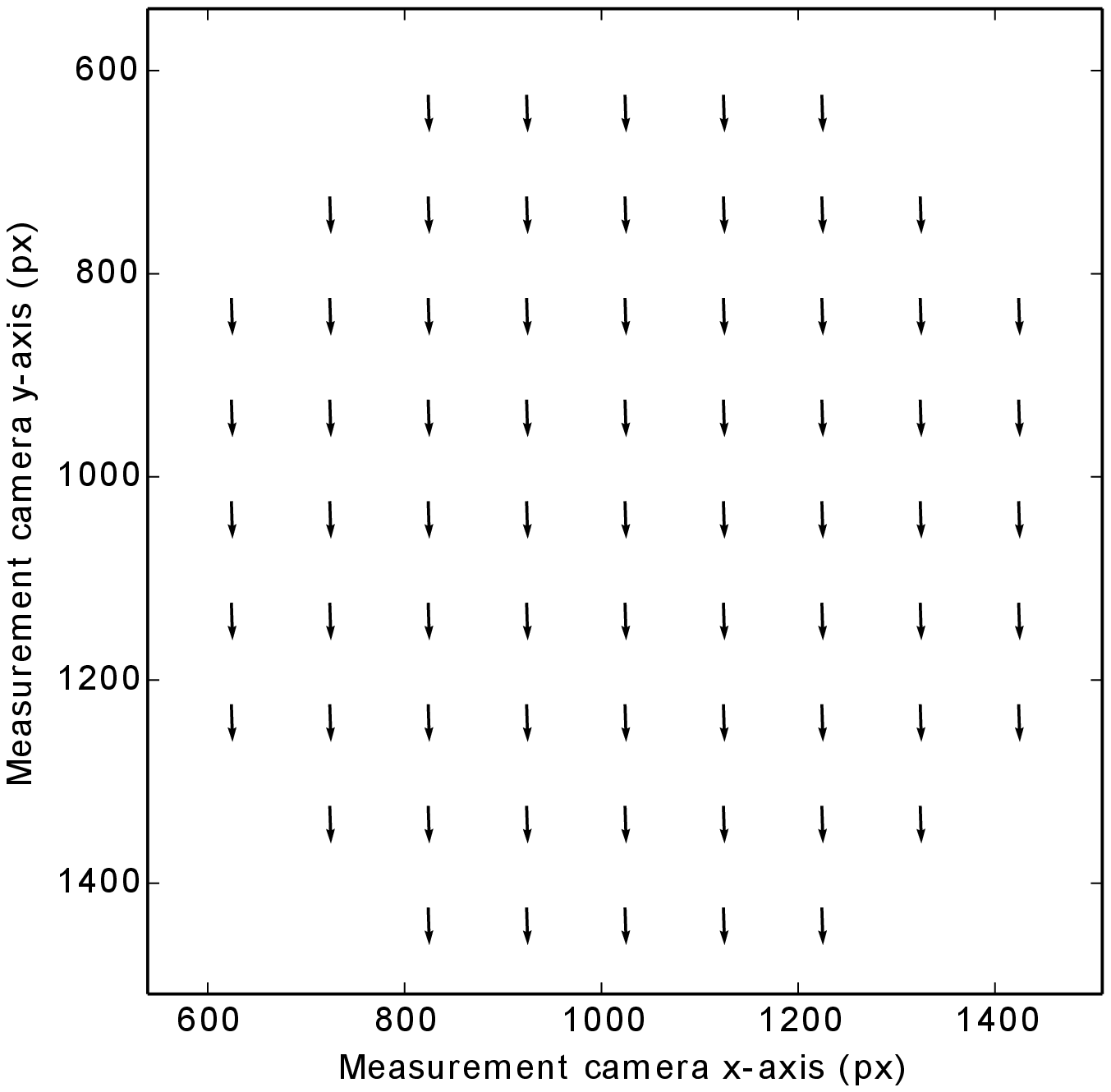}{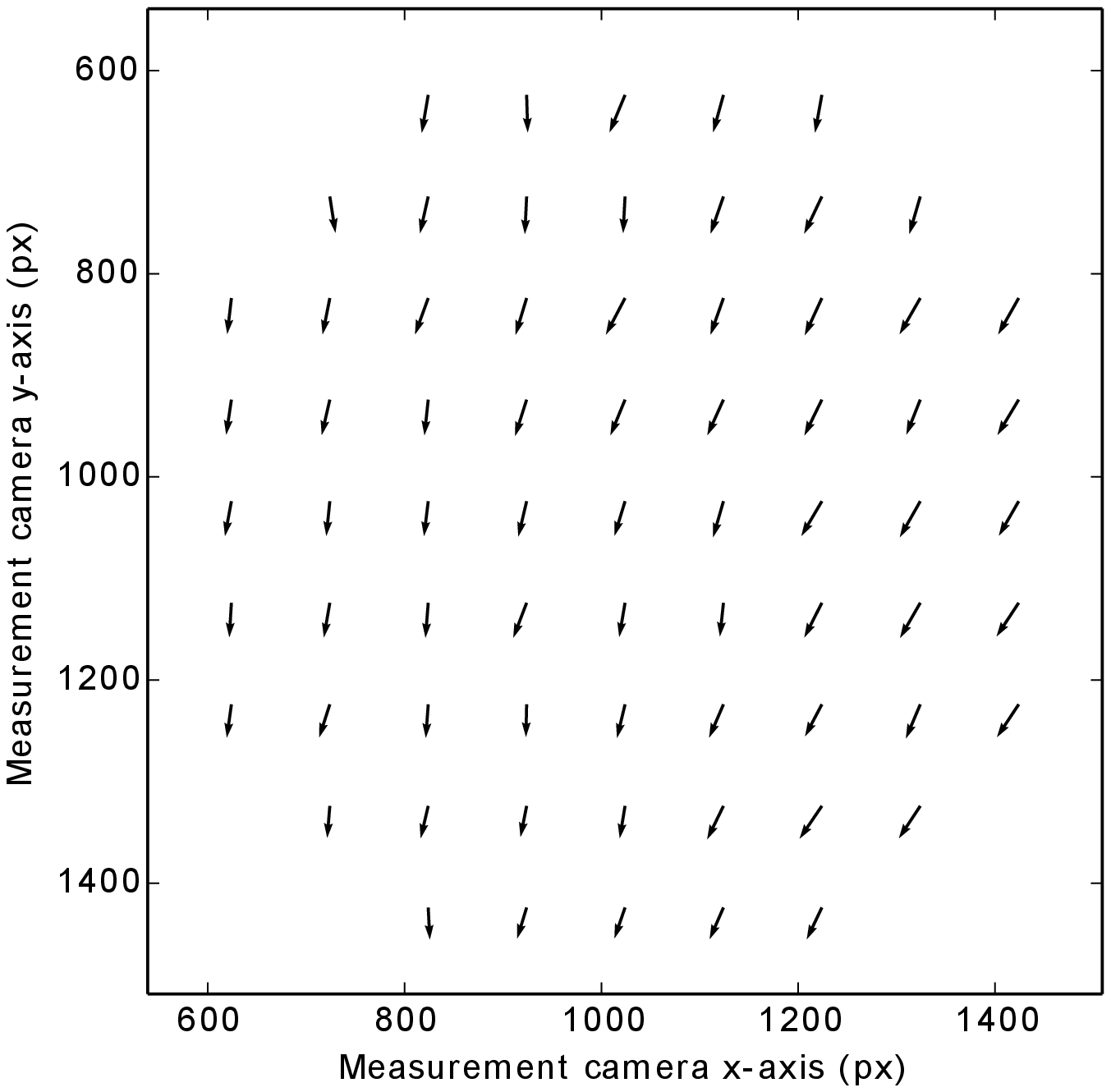}{fig:plot_calGrids}{Calibration vectors for one of six movement directions before (left) and after (right) 1000 positioning cycles with adaptive calibration turned on. Vectors are plotted for every cell in the fiber's 10~mm diameter patrol area, with vector magnitude scaled for visual comparison. This reveals biases as a function of position.}

\section{Conclusions}

The use of a simple recent history of moves, tied to different parts of a fiber's patrol area, has been demonstrated as an effective way to achieve and maintain better positioning performance than with a single global calibration of the spine motor.

A future task is to better understand the effect of the size of the FIFO stack and the number of calibration cells used, although this is likely to be specific to the chosen design of the spine motor.  Indeed, different designs will benefit from this type of adaptive calibration to different extents.

\end{document}